\documentclass[twoside]{article}
\usepackage{fleqn,espcrc2,amssymb,epsf,slashed}
\newcommand{\noi}{\noindent}

\newcommand{\vs}{\vspace}
\newcommand{\hs}{\hspace}
\newcommand{\mquad}{\hspace*{0.25cm}}
\newcommand{\eq}{\begin{equation}}
\newcommand{\en}{\end{equation}}
\newcommand{\eqa}{\begin{eqnarray}}
\newcommand{\ena}{\end{eqnarray}}

\newcommand{\AmS}{{\protect\the\textfont2
  A\kern-.1667em\lower.5ex\hbox{M}\kern-.125emS}}

\title{
{
\vspace{-1.0cm}
\scriptsize
\hfill
\parbox{30mm}{
DESY 99-140\\HLRZ1999\_42
             }
}\\[5mm]
  On the eta--invariant in the four dimensional chiral U(1) theory
}

\author{
  V. Bornyakov\address{\sl Institute for High Energy Physics IHEP, 142284 
    Protvino, Russia},
  A. Hoferichter\address{\sl Deutsches Elektronen-Synchrotron 
                             DESY and NIC, 15735 Zeuthen, Germany} 
  \thanks{Talk given by A. Hoferichter.},
  G. Schierholz$\mbox{}^{\mbox{\scriptsize b,}}$\address{\sl 
  Deutsches Elektronen-Synchrotron DESY, 22603
  Hamburg, Germany},
  A. Thimm\address{\sl Institut f\"ur Theoretische Physik, Freie Universit\"at
    Berlin, 14295 Berlin, Germany}
  }
       
\begin{document}

\begin{abstract}
The imaginary part of the effective action
is investigated in the 4D chiral U(1) theory using the CFA.
\end{abstract}

\maketitle

\section{INTRODUCTION}
%=====================
In the continuum the relation between chiral gauge theories 
in even dimensions $\,D\,$ and their vectorial
counterparts is known since many years. 
For an external gauge field $\,A\,$ in the topologically trivial 
sector, the chiral fermionic effective action $\,W[A]\,$ 
is given by\footnote{Up to local counter 
terms for the real part.} \cite{Alvarez-Gaume:1985xf} :
\eqa
&{\rm Re}&\!\!\!\!\! \Big (W[A] - W[A_0] \Big ) 
= \frac{1}{2} \Big ( W_{\rm V}[A] - W_{\rm V}[A_0] \Big ) 
\nonumber \\
&{\rm Im}&\!\!\!\!\! \Big (W[A] - W[A_0]\Big ) = \pi\eta 
+ 2\pi Q_5^0(A_t) \,,
\nonumber 
\label{main0}
\ena   
\noi
where $\,A_t = (1-t) A_0 + t A\,, \; t\in [0,1],\,$ interpolates between some
initial configuration $\, A_0\,$ and $\,A\,$ in the same topological sector
and $\,W_{\rm V}\,$ denotes the effective action for the associated vector
theory. 
The imaginary part, which conveys the chiral nature of 
the theory, is given by the eta-invariant $\,\eta \,$ \cite{Atiyah} 
and the Chern-Simons form  $\,Q_5^0$.
The latter encodes the anomaly $\,\delta_{g} W[A]\, \propto \,
{\rm i}\, \delta_{g}Q_5^0$, $\,\delta_g\,$ being 
the variation w.r.t. a gauge transformation $\,g$. 
In the anomaly free model the imaginary part is given by the 
eta-invariant alone, manifesting the key r\^ole of $\,\eta\,$ in the 
understanding of chiral gauge theories. 
Since $\,\eta\,$ is defined as the spectral asymmetry of an 
`extended' Dirac operator, its presence indicates a $\,D+1\,$ dimensional
underlying problem. 

%\vs*{0.15cm}
\noi
Despite the recent progress in constructing chiral gauge theories 
on the lattice \cite{chiral_progress}, there seems to be no 
easily manageable expression for $\,\eta\,$ at finite lattice spacing $\,a$. 
Within a 5-dimensional approach a lattice definition of the eta-invariant
(and $\,Q_5^0$) has been given in \cite{aoyama} and shown, that in the 
classical continuum limit the lattice expressions transform to the known
continuum expressions.
From the point of view of a {\sl practical} implementation of chiral 
gauge theories on the lattice it is desirable to control the imaginary part
of the resulting chiral effective action. As a first step towards this
goal we investigate the effective action of the 4D lattice 
chiral U(1) theory in order to extract information on the eta-invariant.  

\section{LATTICE  MODEL AND STRATEGY}
%===================================
In matrix notation with all indices suppressed 
the fermionic action of the model under consideration is 
given by\footnote{We use standard lattice notation for one flavor.}
$$
{S_F}[{\bar \psi}, \psi,U]
 =  %\sum_{x,y}
\,{\bar \psi} \, %_{x}  
{\bf \mathbb{M}}[U]%_{x,y}[U] 
\, \psi%_{y},
$$
\noi 
where $\,U\,$ is an external field with (compact) link variables 
$ U_{\mu}(x) \in \, {\rm U(1)}$ and 
$\,
\mathbb{M}[U] = {\slashed{\mathbb{D}}}[U] + \mathbb{W}[U]\,.
$
The Dirac part is written as
\eqa
\slashed{\mathbb{D}}= \slashed{D}P_{R} + \slashed{D}P_{L} &=& 
\left(
\begin{array}{cc}
                   0        &        {D}_{{L}\,\;~} \\
                   {D}_{{R}\;\,~}        &        0 \\

\end{array} \right)
\nonumber
\ena
\noi
with $\,P_{R,L} = \frac{1}{2} (1 \pm \gamma_5)$, and the Wilson term,

\eqa
\mathbb{W} &=& \left(
\begin{array}{cc}
                   W_{{L}{R}}   &          0            \\
                     0      &        W_{{R}{L}}         \\
\end{array}
\right)   \,,
\nonumber
\ena

\noi
may in principle depend also on $\,U$. Here we will consider 
the so-called ungauged Wilson term only.
In this case there are no counter terms for the imaginary part 
of the effective action \cite{bodwin}.
Explicitely, we consider
\eqa
\mathbb{W} &=& -\frac{1}{2}\left(
\begin{array}{cc}
                   \partial^b\partial^f   &          0            \\
                     0      &        \partial^f\partial^b         \\
\end{array}
\right)\,,   
\nonumber
\ena

\noi
where $\partial^{f,b}\,$ denotes the lattice forward and backward
derivative, respectively.
By choosing the right-handed component of the field $U_R$ to 
be trivial, we obtain a chiral U(1) model with $\,{D}_{R} = 
{\partial}_R$.
In this study, we restrict ourselves to the discussion of 
the imaginary part of the fermionic effective action

$$
{\rm Im}\,W = {\rm Im} \, ( \, -\ln \det \mathbb{M}\,)\,,
$$
 
\noi
which we evaluate in the continuum 
fermion approach (CFA) (e.g.\cite{cfa,bodwin,hernandez}) by the 
following steps.

\begin{itemize}
\item[(1)] interpolate the lattice gauge configuration $\, U^a\,$ 
  on the original lattice with spacing $a$ to a continuum gauge field $\,A$
  by a suitable procedure 
  (e.g. \cite{interpolate}) : $\,U^a \, \longrightarrow \, A$
\item[(2)] in order to avoid infinities, project back 
  $\,A\,$ to a lattice configuration with finer spacing 
  $\,a_f=m^{-1}a,\, m > 1\::\,\,A\,\longrightarrow U^{a_f}$
\item[(3)] consider $\lim_{a_f\to 0}\, {\rm Im}W[U^{a_f}]$, with 
  $\,a\,$ fixed, to evaluate the fermionic effective action for the 
  continuum fermions 
\end{itemize}

\noi
To carry out the full program, eventually the limit $\,a\rightarrow 0\,$
has to be taken, which we will not perform here.
In practice, steps (1) and (2) are combined without making the 
`detour' via the continuum gauge field, such that one interpolates
$\, U^a \rightarrow U^{a_f}\,$ directly. 
We make use of the remaining freedom in the interpolation procedure
by introducing displacement vectors w.r.t. the origins of the original
and finer lattices \cite{axel}. 
Thus, for one configuration $\,U^a\,$ on the original lattice
we are able to obtain a set of configurations $\,\{U^{a_f}\}$.
Corresponding distributions of $\,{\rm Im}W[U^{a_f}]\,$ are used to 
define a variance. In the $\,a_f \to 0\,$ limit the results
should become independent of the details of interpolation, leading to 
narrower distributions for smaller $\,a_f$. 
To compute $\,\det \mathbb{M}[U^{a_f}]\,$
%In step (3) 
we apply a non-Hermitean Lanczos procedure, which has been 
modified for numerical stability  with complete re-orthogonalization. 
%
%\noi
In order to derive
the information about the anomaly free model, we impose 
the anomaly cancellation condition  
$$\,\sum_{f=1}^N \varepsilon_f c_f^3 =0\,,$$
\noi
in an additional step. The sum extends over the $\,N\,$ different 
flavors with chiralities $\,\varepsilon_f\,$ and fermion 
charges $\, c_f$. After anomaly cancellation, we {\sl adopt} the continuum 
notation by setting 

$$
\lim_{a_f\to0}\,{\rm Im} W[U^{a_f}] \, \stackrel{!}{=}\, \pi\eta\,;
\quad{\tiny a={\rm const}}\,,
$$

\noi
which formally provides us with the notion of a lattice eta-invariant. 
Lattice artefacts and gauge invariance breaking effects are expected 
to vanish at least as $\,O(a_f/a)\,$ \cite{bodwin,hernandez}. 

\section{FIRST RESULTS}
%======================
We considered three original\footnote{We set a=1 from now on and
drop the superscript $\,a_f\,$ on interpolated configurations.} 
$\,L^4=3^4\,$ gauge 
configurations denoted by $\,U_{1,2,3}\,$ interpolated to 
lattices sizes up to $\,8^4\,$ as depicted schematically below:

\hs*{2.9cm}$
\{U\}^{L=3}
$

\vs*{-0.05cm}
\hs*{3.0cm} ${\Downarrow}$

\vs*{-0.3cm}
\begin{center}
{\sl random displacements of origin} 
\end{center}

\vs*{-0.3cm}
$ 
\hs*{0.2cm}\Downarrow\hs*{1.2cm}\Downarrow\hs*{1.24cm}\Downarrow
\hs*{1.25cm}\Downarrow\hs*{1.2cm}\Downarrow
$
 
$\{U\}^{L=4}\mquad \{U\}^{L=5} \mquad 
\{U\}^{L=6} \mquad \{U\}^{L=7} \mquad \{U\}^{L=8}\,.$

\begin{figure}[hbt]
%\vspace*{-1.5cm}
\vspace*{-0.7cm}
\begin{center}
%\hspace*{0.27cm}
\hbox{
\epsfysize=9.45cm
\epsfxsize=7.5cm\epsfbox{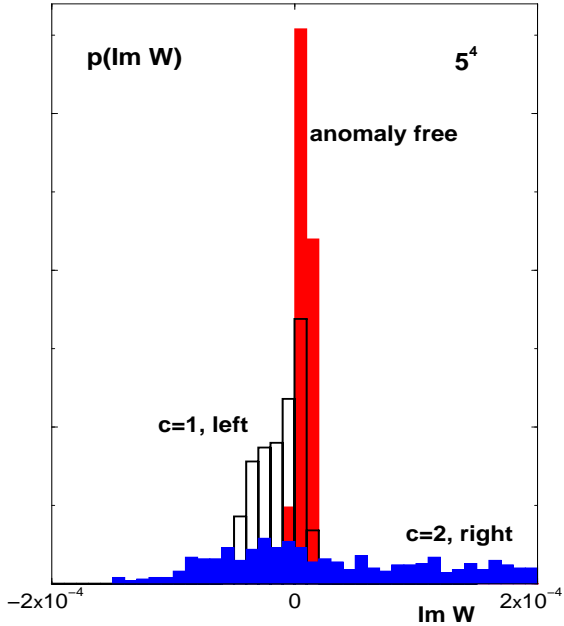}
     }
\end{center}
\vspace*{-1.8cm}
\caption{%\scriptsize
%\small
Distribution of $\,{\rm Im}W\,$ for right (left) handed fermions with 
given charges and for the anomaly free model for $\, U_1\,$ interpolated 
to $\,5^4$. 
}
\vs*{-0.8cm}
%\vs*{-1.4cm}
\label{fig:one}
\end{figure}

\begin{figure}[hbt]
%\vspace*{-0.7cm}
\vspace*{-1.2cm}
\begin{center}
\hspace*{0.27cm}
\hbox{
\epsfysize=9.45cm
\epsfxsize=7.5cm\epsfbox{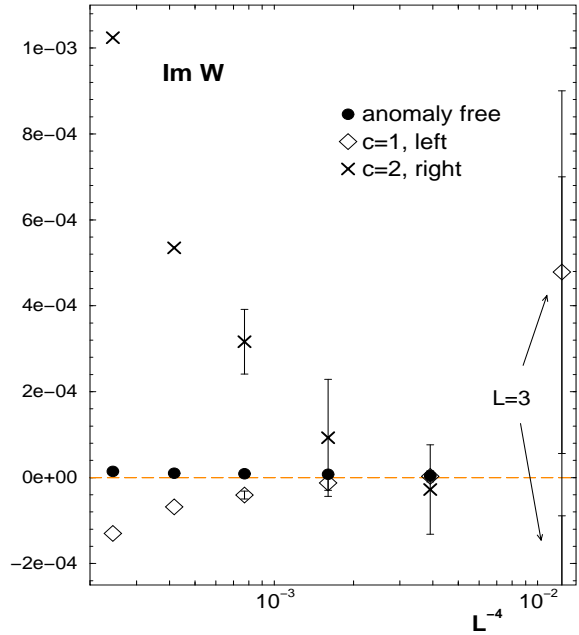}
     }
\end{center}
\vspace*{-1.5cm}
\caption{%\scriptsize 
%\small
${\rm Im}W\,$ vs. $\,L^{-4}\,$ for different charges and chiralities.
The original configuration is $\,U_1\,$.
}
\label{fig:two}
\end{figure}

\vs*{0.2cm}
\noi
Fig.\ref{fig:one} displays the distribution of $\, {\rm Im}W\,$ 
for the case of right (left) handed fermions and after
anomaly cancellation. The original configuration $\,U_1\,$ has been 
chosen randomly, with the constraint $|\,\theta_{\mu}(x)\,| < \frac{\pi}{5}$
on the link angles $\,\theta_{\mu}(x)$. Although the distribution for
the c=2 case has a large variance, after anomaly
cancellation the distribution is sharply peaked about some central 
value. This is possible since the anomaly is canceled event by event, and 
the error bars in Fig.\ref{fig:two}, should not be confused
with Monte Carlo errors, but are an estimation of lattice errors. 
%
% Fig. one
%
%
\noi
The $\,a_f\,$ dependence of $\,{\rm Im} W\,$ for $\,U_1\,$
is shown in Fig.\ref{fig:two}. For the two finest lattices we do not display
error bars, since we could not evaluate the effective action for enough
displacement vectors yet. The observation of Fig.\ref{fig:one} 
continues in the case of finer lattices. In the extreme case, $\,{\rm Im}W\,$
(including the anomaly) is about two orders of magnitude larger than 
the result for the anomaly free model.
For the configurations $\,U_{k}$, $k$=2,3, with
$|\,\theta_{\mu}(x)\,| < \frac{\pi}{6},\,\frac{\pi}{8}$, respectively,
we find similarly small values of $\,{\rm Im}W\,$ in the anomaly free 
case. 

\vs*{-0.2cm}
\section{SUMMARY}
%================
After anomaly cancellation we find  
$\,|\,{\rm Im}W\,| \,<\, 2\cdot 10^{-5}\,$ for the given
configurations. Hence, in these cases, $\,\eta\,$ seems to be 
very small, although it can take any value $\, {\rm mod}\; 2\mathbb{Z}$. 
The results encourage further investigation towards a 
dynamical simulation of a 4D chiral model.
\section{ACKNOWLEDGMENTS}
%=========================
This work has been partially supported by the INTAS 96-370 grant. 
V.B. acknowledges support from RFBR 99-01230a grant.
The calculations have been partly done on the T3E at ZIB, and we
thank H. St\"uben for technical support.
Thanks go to K. Jansen (especially for pointing out the problem of counter 
terms for $\,{\rm Im}W[U]\,$) and to B. Andreas for discussion.
%
% Fig. two
%
%
\vs*{-1.0cm}

\end{document}